\documentclass{article}

\usepackage{arxiv}

\usepackage[utf8]{inputenc} 
\usepackage[T1]{fontenc}    
\usepackage{hyperref}       
\usepackage{url}            
\usepackage{booktabs}       
\usepackage{amsfonts}       
\usepackage{nicefrac}       
\usepackage{microtype}      
\usepackage{graphicx}
\usepackage{natbib}
\usepackage{doi}
\usepackage{wrapfig}

\title{deepbet: Fast brain extraction of T1-weighted MRI using Convolutional Neural Networks}

\author{{Lukas Fisch\textsuperscript{1}} \\
 	\And
	{Stefan Zumdick\textsuperscript{1}} \\
  	\And
	{Carlotta Barkhau\textsuperscript{1}} \\
 	\And
	{Daniel Emden\textsuperscript{1}} \\
 	\And
	{Jan Ernsting\textsuperscript{1,2}} \\
  	\And
	{Ramona Leenings\textsuperscript{1}} \\
   	\And
	{Kelvin Sarink\textsuperscript{1}} \\
   	\And
	{Nils R.~Winter\textsuperscript{1}} \\
   	\And
	{Benjamin Risse\textsuperscript{2}} \\
   	\And
	{Udo Dannlowski\textsuperscript{1}} \\
   	\And
	{Tim Hahn\textsuperscript{1}} \\
   	\AND
	{{\normalfont\textsuperscript{1}University of Münster, Institute for Translational Psychiatry, Germany}} \\
   	\and
	{\textsuperscript{2}Department of Mathematics and Computer Science, University of Münster, Germany} \\
}

\date{}

\renewcommand{\headeright}{}
\renewcommand{\undertitle}{}
\renewcommand{\shorttitle}{deepbet: Fast brain extraction of T1-weighted MRI using Convolutional Neural Networks}

\hypersetup{
pdftitle={deepbet: Fast brain extraction of T1-weighted MRI using Convolutional Neural Networks},
pdfsubject={q-bio.NC, q-bio.QM},
pdfauthor={Lukas Fisch, Stefan Zumdick, Carlotta Barkhau, Daniel Emden, Jan Ernsting, Ramona Leenings, Kelvin Sarink, Nils R. Winter, Udo Dannlowski, Tim Hahn},
pdfkeywords={Brain extraction, Skull stripping, Deep learning, Neural Network},
}

\begin{document}
\maketitle

\begin{abstract}
Brain extraction in magnetic resonance imaging (MRI) data is an important segmentation step in many neuroimaging preprocessing pipelines. Image segmentation is one of the research fields in which deep learning had the biggest impact in recent years enabling high precision segmentation with minimal compute. Consequently, traditional brain extraction methods are now being replaced by deep learning-based methods. Here, we used a unique dataset comprising 568 T1-weighted (T1w) MR images from 191 different studies in combination with cutting edge deep learning methods to build a fast, high-precision brain extraction tool called deepbet. deepbet uses LinkNet, a modern UNet architecture, in a two stage prediction process. This increases its segmentation performance, setting a novel state-of-the-art performance during cross-validation with a median Dice score (DSC) of 99.0\% on unseen datasets, outperforming current state of the art models ($\textsc{DSC} = 97.8\%$ and $\textsc{DSC} = 97.9\%$). While current methods are more sensitive to outliers, resulting in Dice scores as low as 76.5\%, deepbet manages to achieve a Dice score of > 96.9\% for all samples. Finally, our model accelerates brain extraction by a factor of $\approx$10 compared to current methods, enabling the processing of one image in $\approx$2 seconds on low level hardware.
\end{abstract}


\section{Introduction}
The objective of brain extraction is to remove parts of a magnetic resonance imaging (MRI) sample which are non-brain tissue. This process, also known as skull-stripping, stands at the beginning of many popular neuroimage tools such as \citep{ants, afni, freesurfer, cat12, fsl} which preprocess MRIs before further analysis. The quality of brain extraction is critical as any errors during this first preprocessing step can harm the quality of the subsequent preprocessing steps and the downstream analysis.

Established traditional brain extraction tools such as Brain Extraction Tool (BET) \citep{bet}, ROBEX \citep{robex}, BEaST \citep{beast} and 3dSkullStrip, a component of AFNI \citep{afni}, build on specifically handcrafted algorithms which use deformable meshes, prior probabilities and thresholding to segment brain from non-brain tissue voxels. Since convolutional neural networks (CNN) showed their superior performance for image classification \citep{imagenet}, handcrafted pattern recognition algorithms are being replaced with machine learning-based approaches. As a specific subset of pattern recognition, image segmentation was disrupted by the UNet \citep{unet}, a neural network architecture which translated the CNNs superior performance in image classification to image segmentation. HD-BET \citep{hdbet} and SynthStrip \citep{synthstrip} applied UNets with three dimensional (3D) kernels to brain extraction and showed superior segmentation performance compared to the traditional brain extraction tools across modalities.

Here, we aim at further maximizing the segmentation performance by 1. being specific to the most common modality i.e. T1-weighted (T1w) MRI scans of healthy adults; 2. maximizing the amount of different scanners and scanner protocols in the training and testing data to ensure maximum generalizability and stability; 3. using a modern adaptation of the UNet - i.e. LinkNet \citep{linknet} - to enable fast image segmentation while not sacrificing accuracy.

\paragraph{Modal specificity}
Models trained and tested on one specific modality will most likely outperform models which are trained and tested with a wide range of image modalities because the model can “concentrate” on one modality and does not have to co-model other modalities. While our approach can easily be expanded to other modalities, here, we will focus on T1w MRI scans of healthy adults.

\paragraph{Heterogeneous data}
Neural networks learn to recognize the patterns presented during model training. While they are exceptionally good at interpolating between seen data points to adapt to new samples, extrapolation - i.e., generalizing beyond the seen data distribution, posits a challenge. Therefore, the training data should include many cases which represent the edge of the data distribution. In the case of magnetic resonance imaging, these edge cases are mostly images which are subject to intense scanner artifacts and uncommon scanner protocols. Some of these edge cases can be artificially introduced with data augmentation transformations (see Section \ref{sec:augments}). However, one cannot anticipate all edge cases of the final use case and therefore we utilize a heterogeneous pool of training data that include 568 images from 191 different datasets (see Section \ref{sec:augments}) published in OpenNeuro \citep{openneuro}.

\paragraph{Modern CNNs}
Since the proposition of the UNet in 2015, many tricks and trimmings enabled successively higher performing segmentation models while minimizing their computational costs. LinkNet is one descendant of the UNet architecture which optimized the linking between encoder and decoder blocks such that the number of parameters in the network could be reduced, resulting in faster computation \citep{linknet}. The original LinkNet enabled the efficient application of segmentation models to 2D images using 2D kernels. Matching the given 3D MRI data, we utilize a LinkNet with 3D kernels (see Section \ref{sec:augments}).

On top of the 3D LinkNet, we also employ an approach which utilizes 2D CNNs on slices of the original 3D MR image. This method is a popular alternative to the 3D approach \citep{quicknat, fastsurfer} since 3D UNets need much more memory such that high-end GPUs are needed to train them with full-view MR images. Beside the lower memory requirements of 2D CNNs, there exist way more model architectures specialized for 2D images and many of them are available with pretrained parameters. To investigate this issue, we apply an approach using a 2D LinkNet pretrained on ImageNet, which we will call “deepbet 2D”, parallel to the 3D approach, called “deepbet 3D” hereafter. Similar to \citep{fastsurfer} we use adjacent slices as model input and multi-view aggregation - also known as 2.5D approach \citep{han21} - to recapture spatial information in the third dimension. To further smoothen the prediction along the third dimension, we develop a data augmentation technique which interpolates across slices during training (see Section \ref{sec:augments}) and we combine multi-view aggregation with multi-slice aggregation (see Section \ref{sec:procedure}).

All our approaches are two staged: First, the full-view MR image is cropped to the region of interest using a preliminary mask predicted in the first stage. Second, the final mask is calculated, using the cropped MR image as input.

We validate the performance of our approaches using the Dice score metric measured during 5-fold group cross-validation using N = 568 samples from 191 different OpenNeuro datasets. The group cross-validation guarantees that all samples in the validation folds stem from datasets unseen during training of the respective model, resulting in realistic validation performance measures. The results are compared to HD-BET and SynthStrip being the two state-of-the-art deep learning brain extraction models. Finally, we investigate the limitations of each model by visually inspecting the brain masks for the most challenging samples.

\section{Materials and Methods}
\subsection{Datasets}
\label{sec:datasets}
This study utilizes existing data from 191 studies published on the OpenNeuro platform \citep{openneuro}. Data availability is governed by the respective consortia. No new data was acquired for this study.

\begin{figure}
	\centering
	\includegraphics[width=\textwidth]{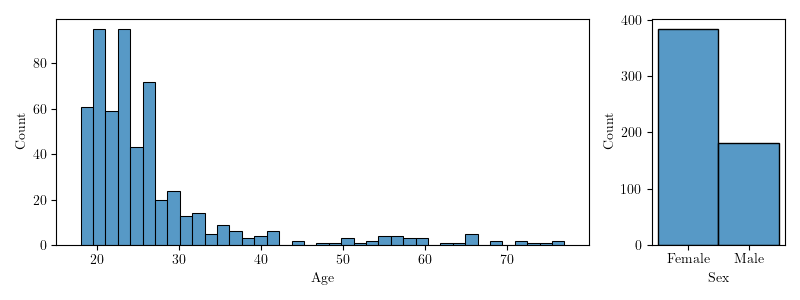}
	\caption{Age and sex distribution across all 568 utilized samples.}
	\label{fig:dataset}
\end{figure}

Out of the 750+ datasets available at OpenNeuro, initially each dataset was included which contained at least five T1-weighted images from at least five adult healthy controls. The samples were preprocessed using the commonly used CAT12 toolbox (built 1450 with SPM12 version 7487 and Matlab 2019a; http://dbm.neuro.uni-jena.de/cat) with default parameters \citep{cat12}. For each dataset the samples were ranked according to the weighted average of the image and preprocessing quality provided by the CAT12 toolbox and the top three samples which passed a visual quality check were finally included. This resulted in a total of 568 samples from 191 datasets (see Figure \ref{fig:dataset}).

By including only the three images of each dataset which show the highest preprocessing quality, we maintain a high quality standard regarding the CAT12 tissue segmentation masks and consequently the ground truth masks we will use for training and validation (see Section \ref{sec:masks}). This way we avoid dealing with “wrong” ground truth masks which would complicate the qualitative analysis (see Section \ref{sec:analysis}).

The models are trained using all 568 samples from 191 studies using 5-fold group cross-validation based on their OpenNeuro dataset identifier. This way all samples in the validation folds stem from datasets unseen during training of the respective model resulting in realistic validation performance measures.


\subsection{Ground Truth masks}
\label{sec:masks}
The ground truth masks are derived based on tissue segmentation masks generated by the CAT12 toolbox which have been meticulously quality checked (see Section \ref{sec:datasets}). These tissue segmentation masks contain the probability for the background and the foreground tissue classes cerebrospinal fluid, grey matter and white matter in each voxel. Brain extraction masks can be easily derived by summing up the probability values of all foreground classes. We directly use this probability mask and initially do not apply thresholding (e.g. $p_{foreground} < 0.5 \rightarrow 0$, $p_{foreground} \geq 0.5 \rightarrow 1$) as this would omit the measure of uncertainty CAT12 generates.

\subsection{Preprocessing}
\label{sec:preprocessing}
Before model training, the images are preprocessed using bias correction followed by intensity normalization. The ANTS implementation \citep{ants} of the standard n4 bias field correction \citep{n4_bias_corr} is applied. Then the intensity values of each sample are clipped into a range between the 0.5\% and 99.5\% quantile as proposed in \citep{nnunet}. Finally, the image intensity is normalized to a mean of 0.449 and a standard deviation of 0.229, aligning it to the ImageNet \citep{imagenet} intensity distribution the encoder of the 2D model has been pretrained on (see Section \ref{sec:architecture}).

\subsection{Data Augmentation}
\label{sec:augments}
Data augmentation is used during model training to introduce artificial effects which can occur in potential use cases. This increases model generalizability since effects which are infrequent in the training data can be systematically oversampled with any desired intensity.

\begin{figure}
	\centering
	\includegraphics[width=\textwidth]{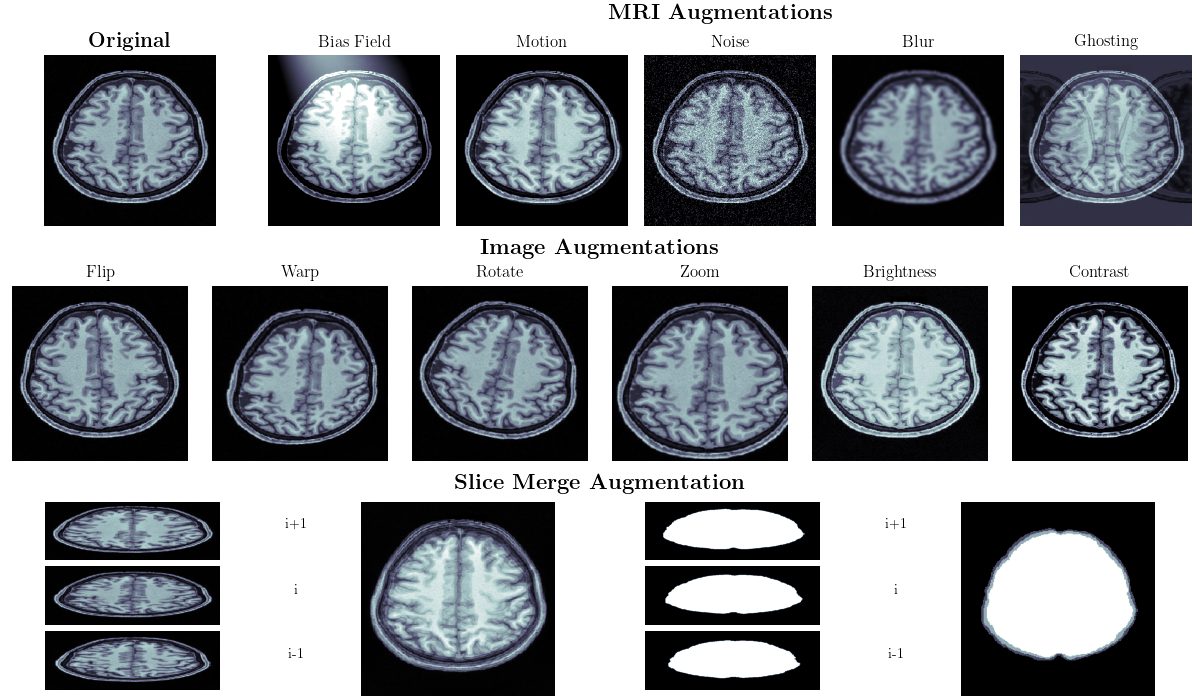}
	\caption{All utilized MRI specific augmentations (top), standard image augmentations (middle) and the slice merge augmentation transformation (bottom), based on an axial slice of an example image (top left).}
	\label{fig:augments}
\end{figure}

We combine standard image augmentations with augmentations specific to magnetic resonance images and augmentation specific to the 2D approach which merges adjacent slices (see Figure \ref{fig:augments}). The standard image augmentations consist of four spatial transformations (flip, warp, rotate and zoom) and two intensity transformations (brightness and contrast). These transformations are implemented using aug\_transforms function of the fastai package \citep{fastai} with max\_rotate set to 15, max\_lighting set to 0.5, max\_warp set to 0.1 and default settings for all remaining parameters.

The five MRI specific augmentations are: Bias fields, motion artifacts, noise, blurring and ghosting. Bias fields are simulated with a linear combination of polynomial basis functions \citep{lee99} with the order 4 and ghosting is achieved by introducing artifacts in the k-space of the image. The implementation of these augmentations is based on the torchio package \citep{torchio}.

Finally, we use a transformation specific to the 2D model (see Section \ref{sec:architecture}) which incorporates the adjacent slices $i+1$ and $i-1$ (see Figure \ref{fig:augments} bottom) into slice $i$ with $x_i = (1- \alpha) * x_i + \alpha * (x_{i+1} + x_{i-1})$. Alpha is randomly sampled between 0 and 0.5 and the transformation is applied to the image slice and the respective mask slice. This augmentation aims to reinforce the consistency of the predicted masks along the third dimension such that slicing artifacts (see Figure \ref{fig:prediction}B center) are minimized.

\subsection{Model Architecture}
\label{sec:architecture}
\paragraph{2D Model}
The 2D model is a two dimensional (2D) convolutional neural network (CNN) which is using five neighboring slices to predict the segmentation mask of the central of these five slices. The 2D CNN is a LinkNet \citep{linknet} which is an advancement on the U-Net architecture specialized to produce accurate segmentation masks with high efficiency. To utilize transfer learning, a 2D CNN pretrained on the ImageNet dataset was used as the encoder part of the LinkNet. The encoder's architecture is a GENet (GPU-Efficient Network) \citep{gernet} with five input channels, corresponding to the five neighboring coronal slices which are inputted to the network. The model is implemented using the Segmentation Model PyTorch package \citep{segmentation_models_pytorch} with default parameter setting except for the encoder depth which is set to 4 \citep{nnunet}.

\paragraph{3D Model}
The 3D model is a 3D LinkNet which is based on the original 2D implementation (https://github.com/e-lab/pytorch-linknet) employing 3D convolution and 3D pooling instead of the 2D operations. Since the 3D MR images need more memory compared to 2D images, two additional modifications had to be implemented: First, dividing the number of channels used for each convolutional operation by 4 reducing the needed memory. Second, the batch normalizations \citep{batchnorm} were replaced by Instance Normalizations \citep{instancenorm} such that model training could be done with a batch size of 1.

\subsection{Training Procedure}
Both models are trained with a learning rate of .001 using a combination of the Dice Loss and the Focal Loss using the GeneralizedDiceFocalLoss of the monai package \citep{monai} with lambda\_focal set to 0.2. The 2D model is trained with a batch size of 32 while the 3D model is trained with a batch size of 1 due to memory constraints. To equalize the amount of training, i.e. the total number of batch iterations, the three 2D models (sagittal, coronal and axial) are each trained for 10 epochs while the 3D model is trained for 200 epochs resulting in $\approx$100.000 total iterations for both approaches, respectively.

Three methods for quick convergence are applied: usage of the Ranger optimizer, the Flatten + Cosine Annealing learning rate schedule and Discriminative learning rates. The Ranger optimizer proposed by \citep{ranger21} combines the rectified Adam \citep{liu21} optimizer which increases stability during the beginning of the training with LookAhead \citep{lookahead} which speeds up convergence during the end of the training. The Flatten + Cosine Annealing learning rate schedule was specifically developed for the Ranger optimizer combining constant learning rates for stable exploration with cosine annealing to smoothly finish training. To avoid catastrophic forgetting in the 2D model the learning rate of pretrained model encoder is set to zero while the decoder and the head are trained with 20\% and 100\% of the scheduled learning rate, respectively. These methods are implemented using the fastai package with the respective default parameters.

\subsection{Prediciton Procedure}
\label{sec:procedure}
\begin{wrapfigure}{r}{7.8cm}
	\centering
	\includegraphics[width=6.7cm]{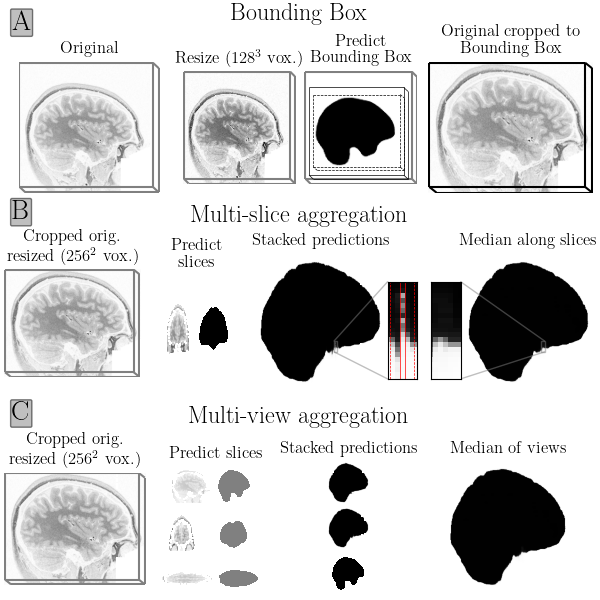}
	\caption{A.) First stage of the prediction process: The original MR image is resized to 128³ voxels, the preliminary mask is predicted and the image is cropped with a margin of 10\% of the respective edge length of the minimal bounding box. B.) Multi-slice aggregation: Five adjacent (coronal) prediction slices are aggregated via the voxelwise median. C.) Multi-view aggregation: The sagittal, coronal and axial view, i.e. the stacked predictions of the respective models, are aggregated via the voxelwise median.}
	\label{fig:prediction}
\end{wrapfigure}

The prediction procedure consists of two successive stages: First, a preliminary mask is predicted using the full-view MRI with a low resolution (128³ voxels) to determine the minimal bounding box which contains all brain voxels in the preliminary mask (see Figure \ref{fig:prediction}A). With a margin of 10\% of the respective edge length, the image is cropped around the minimal bounding box and resampled to 256³ voxels. Then the second stage prediction is done to obtain the final brain mask.

In the second stage prediction we compare the 3D approach with the popular, alternative 2D approach: The 3D approach (called “deepbet 3D”) applies a 3D CNN to the MR image while the 2D approach (called “deepbet 2D”) applies 2D CNNs to individual slices of the MR image (see Section \ref{sec:architecture}).

The final prediction of deepbet 2D utilizes a multi-view aggregation approach i.e. predictions are done on sagittal, coronal and axial slices separately using three models trained on the respective views and then these predictions are aggregated \citep{quicknat}. The aggregation of the three-view predictions are typically done by calculating the voxelwise median to get a smooth ensemble prediction (see Figure \ref{fig:prediction}C). To further minimize slicing artifacts, we combine multi-view aggregation with multi-slice aggregation. Multi-slice aggregation incorporates predictions of neighboring slices into each slice resulting in masks with smoother transitions between slices (see Figure \ref{fig:prediction}B). Here, multi-slice aggregation is done with n=5 slices such that the prediction of slice i is aggregated via the voxelwise median of the slices [i-2;i+2]. To maximize the accuracy of the final prediction we combine multi-view and multi-slice aggregation such that each voxel of the final prediction is based on the median of an ensemble of 15 - i.e. 3 views * 5 slices - voxel predictions.

Finally, two standard postprocessing steps are applied to the predicted mask: Firstly, voxels which do not belong to the largest connected component are set to zero using the cc3d package \citep{cc3d} and secondly, holes in the remaining component are filled using the fill\_voids package \citep{fill_voids}.

\subsection{Evaluation}
\label{sec:evaluation}
To evaluate the agreement between the predicted brain masks and the thresholded probabilistic ground truth the Dice score is calculated. As the probability values of our ground truth masks suggest, there is room for interpretation where to exactly draw the boundary line of the brain mask.

\begin{figure}[h]
	\centering
	\includegraphics[width=7cm]{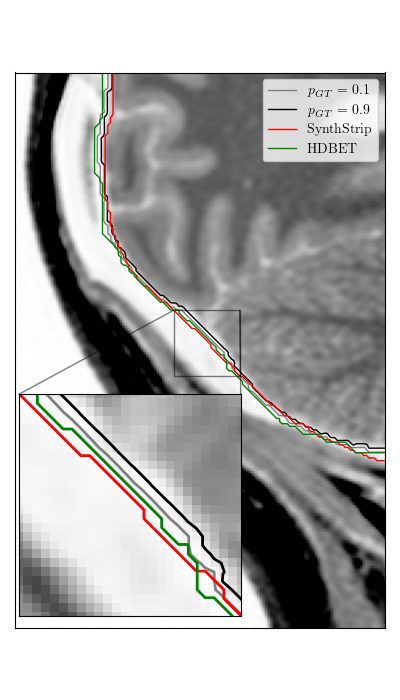}
	\caption{Boundary line of the brain masks generated by SynthStrip, HDBET and the ground truth probability mask thresholded at 0.1 and 0.9.}
	\label{fig:thresholds}
\end{figure}

As shown in Figure \ref{fig:thresholds}, some tools include a thicker layer of cerebrospinal fluid around the brain then others. To account for this, SynthStrips mask border threshold is calibrated to -1 millimetre as this setting resulted in the best agreement to our ground truth masks in terms of the Dice score. Similarly, we calibrate the threshold which is used to binarize the ground truth mask before calculating the Dice score individually for HD-BET and SynthStrip. We find that a threshold of 0.5 for SynthStrip and 0.5 for HD-BET maximizes the median Dice score for each tool, indicating proper calibration.

\newpage

\section{Results}
\begin{figure}[h]
	\centering
	\includegraphics[width=\textwidth]{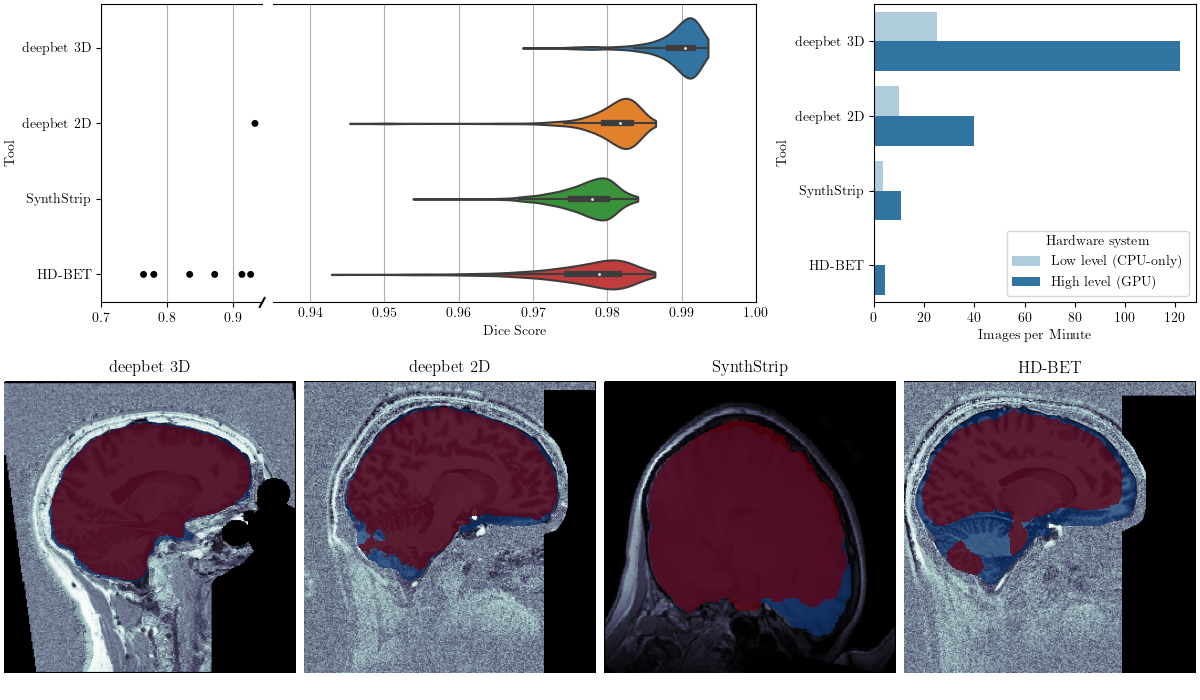}
	\caption{Top: Dice scores and average images per minute processed by deepbet 3D, deepbet 2D, SynthStrip and HD-BET. Bottom: Sagittal slice of the worst predictions of each method (red) and the respective ground truth masks (blue).}
	\label{fig:results}
\end{figure}

During cross-validation deepbet 3D performs best regarding the Dice scores (DSC) across the validation samples (see Figure \ref{fig:results}). It shows the highest median DSC of 99.0\%, followed by deepbet 2D ($\textsc{DSC} = 98.2\%$). SynthStrip ($\textsc{DSC} = 97.8\%$) and HD-BET ($\textsc{DSC} = 97.9\%$) show the lowest median DSC of the investigated methods.

On top of the highest performance, deepbet 3D achieves the highest processing speed on both low level and high level hardware. On low level hardware deepbet 3D takes 25 images per minute which translates into a 2.5x speedup compared to deepbet 2D (10 images per minute), a 7x speedup compared to SynthStrip (3.5 images per minute) and a 250x speedup compared to HD-BET (0.1 images per minute). On high level hardware deepbet 3D achieves 122 images per minute being 3x faster than deepbet 2D (40 images per minute), 11x faster than SynthStrip (11 images per minute) and 27x faster than HD-BET (4.5 images per minute). We use a laptop with a Intel i7-8565U Central Processing Unit (CPU) and without a Graphical Processing Unit (GPU) as low level hardware and a system with a AMD Ryzen 9 5950X CPU and a NVIDIA GeForce RTX 3090 GPU as high level hardware.

\subsection{Qualitative analysis}
\label{sec:analysis}
By examining the most challenging images (see Figure \ref{fig:results} bottom), i.e. the images with the lowest Dice score, deficits of the respective approach can be unveiled. All extraction methods are challenged by either of two issues: First, strong noise occurring in images of the OpenNeuro studies ds001168 and ds003192. Second, atypically strong rotation around the sagittal axis in images of study ds001734.

In deepbet 3D, deepbet 2D and HD-BET the respective images with the lowest Dice score are the noisy images. deepbet 3D maintains the highest minimal DSC of 96.9\%, excluding the outer edge of the cerebellum for one of the noisy images. HD-BET is most sensitive, including large patches of noise in the background into the brain mask, resulting in the strong outliers with Dice scores as low as 76.5\%. deepbet 2D tends to exclude small parts where strong noise of the background blends into the cerebellum from the brain mask resulting in a minimal DSC of 93.4\%. This showcases the drawback of the 2D approach compared to the 3D approach: 3D models can infer the segmentation class of some parts of the image which are “occluded” (e.g., by noise) by memorizing the typical 3D shape of the masks during training which 2D models cannot.

\begin{figure}[h]
	\centering
	\includegraphics[width=\textwidth]{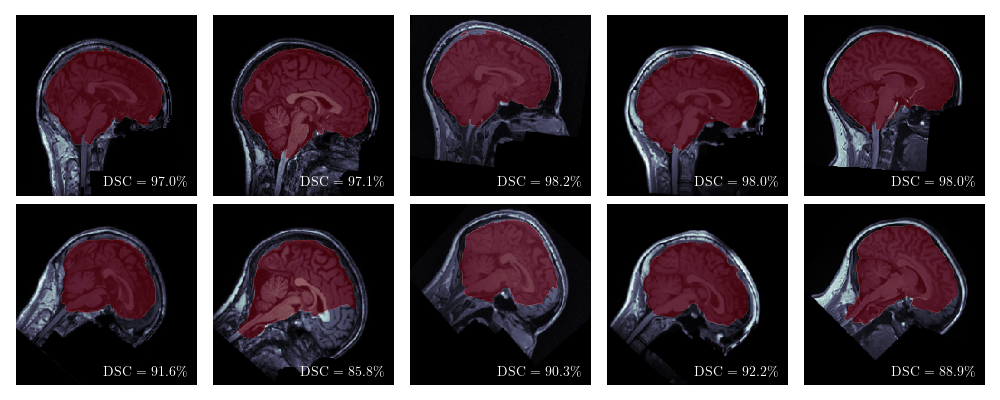}
	\caption{Systematic errors of SynthStrips brain extraction caused by 40 degrees of rotation around the sagittal axis. Five original (top) and rotated (bottom) samples shown with their respective brain mask (red) and Dice score (DSC).}
	\label{fig:tilted}
\end{figure}

With regard to SynthStrip, we noticed that it included small regions of tissue around the cerebellum (see Supplement Figure \ref{fig:worst_synthstrip}) along with the known issues of including tissue around the eye socket and regions of extra-cerebral matter near the dorsal cortex (Hoopes et al., 2022). However, its lowest Dice score of 95.4\% is caused by excluding a large part of the frontal lobe in one of the images which is atypically rotated around the sagittal axis. This behaviour can be replicated in other images by rotating them in the same direction before applying brain extraction (see Figure \ref{fig:tilted}). Despite SynthStrip being trained with rotations between -45 and 45 degrees via data augmentation, the rotation of 40 degrees around the sagittal axis causes it to exclude large parts of the frontal lobe in all five cases. This indicates that the SynthStrip model is overfitted to typically oriented, non-tilted samples.

\section{Discussion}
The results confirm that with deepbet efficient brain extraction of T1w MRIs can be done with high-precision i.e. a median Dice score (DSC) of 99.0\%. deepbet is outperforming the state-of-the-art tools SynthStrip ($\textsc{DSC} = 97.8\%$) and HD-BET ($\textsc{DSC} = 97.9\%$) which achieve higher median Dice scores than in the T1w test images of their respective original studies, indicating proper calibration (see Section 2.8). Furthermore, deepbet manages to maintain high precision even for edge cases (e.g. images with strong background noise) with a minimal DSC of 96.9\% beating SynthStrip and HD-BET with minimal Dice scores of 95.4\% and 76.5\%. On top of that, deepbet is ~10x more efficient than current state-of-the-art tools.

Besides the best performing approach (called deepbet or deepbet 3D), another approach (called deepbet 2D) which applies 2D CNNs on 2D slices of the 3D image was investigated. The results show that this approach is inferior to deepbet 3D since 1. It is susceptible to strong noise occluding parts of the brain and 2. It introduces slicing artifacts which have to be smoothed out by multi-slice and multi-view aggregation (see Section 2.7).

Due to the two staged prediction process and the usage of LinkNet instead of the standard UNet architecture deepbet is highly computationally efficient, such that brain extraction can be done in two seconds on low level hardware (here Intel i7-8565U laptop CPU) and half a second on high level hardware (here AMD Ryzen 9 5950X CPU, NVIDIA GeForce RTX 3090 GPU). This makes deepbet attractive for integration into neuroimaging pipelines and enables processing large datasets within a single day without the need to access large compute clusters. The method is made available at \url{https://github.com/wwu-mmll/deepbet}.

In this work, we purposely limited deepbet to T1w MRIs of healthy adults as it is the dominant modality in the neuroimaging field and modal-specific training and application maximizes performance. In future work, this modal-specific approach can be applied to other modalities (e.g. T2w, FLAIR, DWI, PET, CT) and patient groups (e.g. children and brain-tumor patients) by training the model on respective images, optimally stemming from a large number of different studies.

\section*{Declarations}
\subsection*{Conflict of Interest}
All authors declare that they have no conflicts of interest.

\subsection*{Funding}
This work was funded by the German Research Foundation (DFG grants HA7070/2-2, HA7070/3, HA7070/4 to TH) and the Interdisciplinary Center for Clinical Research (IZKF) of the medical faculty of Münster (grants Dan3/012/17 to UD and MzH 3/020/20 to TH and GMzH).

\bibliographystyle{unsrtnat}
\bibliography{references}  





\newpage

\section*{Supplementary Material}

\setcounter{figure}{0} 
\renewcommand{\thefigure}{S\arabic{figure}}

\begin{figure}[h]
	\centering
	\includegraphics[width=\textwidth]{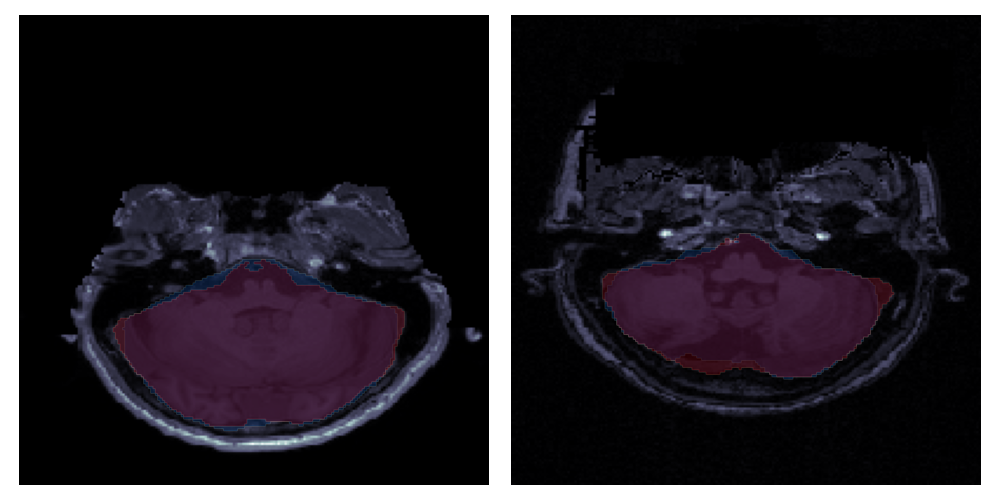}
	\caption{Two example cases in which SynthStrips brain masks (red) includes small regions of tissue around the cerebellum. Ground truth brain mask depicted in blue.}
	\label{fig:worst_synthstrip}
\end{figure}

\end{document}